\documentclass [prl,amsmath,showpacs,twocolumn,preprintnumbers,superscriptaddress]{revtex4-1}

\providecommand{\mbf}{\mathbf}

\usepackage{graphicx}  
\usepackage{dcolumn}   
\usepackage{amsmath}   
\usepackage{bm}        
\usepackage{amssymb}   

\usepackage{mathtools}
\usepackage{amssymb}
\usepackage{bm}
\usepackage{braket}

\RequirePackage{xspace}

\renewcommand*{\vec}[1]{\mbf{#1}}

\newcommand*{\J}{\vec{J}}

\begin{document}

\title{Constraining the Generalized Uncertainty Principle with the light twisted by rotating black holes and M87*}

\author{Fabrizio Tamburini}
\email{fabrizio.tamburini@gmail.com}
\affiliation{ZKM -- Zentrum f\"ur Kunst und Medien,\\ Lorenzstra{\ss}e 19, D-76135 Karlsruhe, Germany}%
\author{Fabiano Feleppa}
\email{feleppa.fabiano@gmail.com}
\affiliation{Institute for Theoretical Physics, Utrecht University, Princetonplein 5, 3584 CC Utrecht, The Netherlands}%
\author{Bo Thid\'e}
\email{bothide@gmail.com}
\affiliation{Swedish Institute of Space Physics, Uppsala, Sweden}%


\begin{abstract}
We test the validity of the Generalized Uncertainty principle in the presence of strong gravitational fields nearby rotating black holes; Heisenberg principle is supposed to require additional correction terms when gravity is taken into account, leading to a more general formulation also known as the Generalized Uncertainty Principle.  
Using as probes electromagnetic waves acquiring orbital angular momentum when lensed by a rotating black hole, we find from numerical simulations a relationship between the spectrum of the orbital angular momentum of light and the corrections needed to formulate the Generalized Uncertainty Principle, here characterized by the rescaled parameter $\beta_0$, a function of the Planck's mass and the bare mass of the black hole.
Then, from the analysis of the observed twisted light due to the gravitational field of the compact object observed in M87, we find new limits for the parameter $\beta_0$. With this method, complementary to black hole shadow circularity analyses, we obtain more precise limits from the experimental data of M87*, confirming the validity of scenarios compatible with General Relativity, within the uncertainties due to the experimental errors present in EHT data and those due to the numerical simulations.
\end{abstract}


\maketitle

\section{\label{sec:level1}Introduction}
Most of the knowledge we have about our Universe is obtained by extracting the information encoded in the spectrum of electromagnetic (EM) waves emitted by celestial bodies and, more recently, from neutrinos and gravitational waves in the framework of multi-messenger astronomy \cite{multi1,multi2,multi3}, including new phenomena observed at extremely high energies and in strong gravitational fields, where the classical formulation of the Heisenberg uncertainty principle (HUP) can loose its validity. Therefore, a modification of the HUP is considered to study physics at very high energies or, equivalently, at short distance scales, leading to the so-called Generalized Uncertainty Principle (GUP); among other things, quantum gravitational corrections are expected to prevent black hole (BH) evaporation near the Planck scale \cite{Chen}, as also suggested in \cite{erepr}.
The formulation of a GUP has implications in various fields \cite{Pan, Moradpour, EPL, Ali, Garattini, Medved, Myung, Park, Nozari1, Majumder, Vagenas, Anacleto, Feng, Das1, Sprenger, Zhu, Das2, Nozari2, Yang, Zhang, Faizal1, Faizal2}; black hole physics, String Theory \cite{string1,string2}, Loop Quantum Gravity \cite{lqg}, Deformed Special Relativity \cite{Ciafaloni1, Gross, Ciafaloni2, Paffuti, Veneziano, Maggiore, Kempf1, Kempf2, Kempf3, Scardigli1, Adler, Scardigli2}, where not only General Relativity (GR) but also Quantum Mechanics (QM) breaks down, suggest that the HUP can be obtained by introducing quantum-gravitational corrections to the classical definition of the commutator of two conjugate variables, such as the coordinate $x$ and the momentum $p$ with their corresponding operators $\hat x$ and $\hat p$, $\left[ x, p \right] = i \hbar$. In order to test the validity of the GUP in strong gravity conditions, we use a new method that is now starting to be considered in astrophysics and could give a hint to multi-messenger astronomy: the exploitation of the whole set of total and partial conserved quantities of the EM field to obtain more information about the source from the EM waves collected during observations \cite{fuschichNikitin1}. Specifically, we will focus on the use of the EM orbital angular momentum (OAM) \cite{Torres&Torner:Book:2011,huptambu} of light lensed by a Kerr BH to constrain the GUP parameter $\beta$. In fact, up to now, astronomical observations have used only a small subset of the EM field properties such as energy, related to the invariance with respect to time, intensity and polarization. Energy, for example, is related to the frequency of the EM wave and through spectroscopy analyses chemical abundances are revealed; including the Doppler effect one gets additional information about the motion of a source or reveals the presence of a gravitational well. Similarly, by using in a clever way the intensity of the field one reveals e.g. the distance of a galaxy with standard candles. Theorized by M. Abraham in $1914$ \cite{abraham}, OAM is one of the two components of the EM total angular momentum $\J$, which is, as is well-known, a conserved quantity of the EM field concomitant with the ten-dimensional Poincar\'e  group of Noether invariants \cite{fuschichNikitin2}. The quantity $\J$ is composed of and transported in two different forms: the spin angular momentum (SAM)  $\vec{S}$, and $\vec{L}$ (OAM), in a superposition $\J=\vec{S}+\vec{L}$; SAM and OAM, however, do not generally behave as two independently separated EM conserved quantities: an interplay between OAM and SAM occurs.
Only in the paraxial approximation OAM and SAM behave as they were two independently propagating gauge--invariant observables. SAM is associated with the polarization of light and its helicity and, down to the quantum level, finds correspondence with the spin of the photon, while OAM is related to the spatial configuration of the beam in amplitude and phase.
Each OAM beam is characterized by a number $\ell$ of twists in its azimuthal phase and $p$ radial modes, and each photon of the beam carries a quantized amount of angular momentum, namely $\pm \ell \hbar$ \cite{laser,Torres&Torner:Book:2011}, down to the quantum level \cite{mair}. OAM found applications in different fields of research and technology, including radio waves \cite{thide2007utilization,tamburini2011} and radio \cite{tamburini2012encoding,spinello2015experimental,triple,someda,oldoni} and optical \cite{huang2014100} telecommunications, far beyond the classical multiplexing schemes \cite{klemes}.

The novel use of the EM OAM offers new tools of investigation for astronomy and astrophysics  \cite{20,27}. 
The most striking result obtained so far is the measure of the rotation parameter of M87$^{*}$ \cite{29,IJMPD,PRA}. 
Rotating BHs, initially described by Kerr \cite{28} and then observed by the Event Horizon (EHT) collaboration \cite{30a,30b,30c,30d,30e,30f}, can be revealed by the presence of twisted light \cite{28b}: when light passes nearby a rotating BH, the geometry of this type of spacetime imparts a twist in light's spatial phase distribution, thus revealing the BH's rotation. 

Also, OAM can be naturally emitted or imposed to the light of distant sources also by other different astrophysical phenomena: EM waves can acquire OAM when they traverse peculiar regions in space containing, e.g., plasma inhomogeneities, in which case they are resonant with the turbulent plasma \cite{27a,27b}. 
The use of OAM in astronomical instrumentation allowed the first direct imaging of extrasolar planets \cite{mawet} by artificially imposing OAM to the light coming from celestial bodies \cite{21,22,23,24,25} and improve the resolving power of any optical instrument up to one order of magnitude \cite{26}.

In this letter we present new limits for the GUP from the OAM analysis of spacetimes  of rotating BHs by using the software KERTAP \cite{kertap}; GUP corrections have been obtained and then compared with the results coming from the OAM analysis of EHT data \cite{29}.

\section{GUP limits from M87* with OAM}

It is usually assumed that, in order to formulate the GUP, one has to modify the HUP by using a dimensionless parameter $\beta$ (with a value which is a priori unknown and thought to be of the order of unity or slightly different \cite{DasVagenas,Das5,Scardigli2}) and including in the formulation the presence of linear and quadratic terms, $[x,p]= i \hbar (1-2 \beta p + \beta^2 p^2)$, leading to
\begin{equation}\label{gup}
\Delta x \Delta p \geq \frac{\hbar}{2} \left(1 + \beta \frac{l^2_p}{\hbar^2} \frac{}{} \Delta p^2 \right),
\end{equation}
when we consider e.g. the position, $x$, and the conjugate momentum, $p$, of a test particle with their corresponding quantum observables, $\hat x$ and $\hat p$; in Eq. (\ref{gup}), $l_p$ refers to the Planck length.

However, the choice of $\beta \sim 1$ renders quantum gravity effects too small to be measured; therefore, without imposing a priori any constraint on the value of the GUP parameter $\beta$, current experiments can predict larger upper bounds on it, which are compatible with observations \cite{DasVagenas, Feleppa, Das4, Marin1, Das5, Marin2, Bawaj, Casadio, Gao, Yunes, Bushev, Yang2, ghosh}, including extreme astrophysical scenarios like those that can be observed in the neighborhoods of black holes or with alternative theories of gravity in the presence of dilaton BHs \cite{mizuno} or boson stars \cite{olivares}. 

A new quantum-corrected Schwarzschild solution, recently proposed in \cite{Carr2015}, connects the deformation of the Schwarzschild metric of a static BH directly to the GUP uncertainty relation, without relying on a specific representation of commutators. Moreover, such a solution has been extended to charged and rotating black holes in Ref. \cite{Carr2020}, followed by the analysis of the M$87^{*}$ BH's shadow \cite{Neves,Jusufi}.
In the case of rotating BHs, an effect of the GUP can be found in the variation of the Arnowitt-Deser-Misner (ADM) mass of a BH. 

Let us now define the parameter $\beta_0= \beta / 2M^2$ (the rescaled GUP parameter). The ADM mass of the black hole is modified as $M' = M + \beta_0 M_{p}^2$, where $M$ is its bare mass and $M_p$ denotes the Planck mass. 
For any given value of the GUP parameter $\beta$, there should exists a critical spin above which the solutions bifurcate into sub-Planckian and super-Planckian phases, separated by a mass gap in which no black holes can form.
Hereinafter we will use Planck's units, where $M_p$ is set to unity as well as the speed of light $c$, the Planck constant $\hbar$ and the gravitational constant $G$.

The quantum version of a rotating BH can be obtained from the line element of the Kerr metric with mass $M$ and angular momentum $J$  \cite{Carr2020}. In Boyer-Lindquist coordinates,
\begin{align}
ds^2 &= -Adt^2 + \frac{\rho^2}{\Delta}dr^2 - \frac{2r_S r a \sin^2\theta}{\rho^2}dtd\phi + \rho^2 d\theta^2 \nonumber \\
&\hspace{0.4cm} + \left(r^2 + a^2 + \frac{r_S r a^2}{\rho^2}\sin^2\theta\right)\sin^2\theta d\phi^2,
\end{align}
where $r$ is the spheroidal radial coordinate, $r_S = 2M$ the gravitational radius, $A = 1- r_S r / \rho^2$, $\rho^2 = r^2 + a^2 \cos^2\theta$ and $\Delta = r^2 - r_S r + a^2$ are the quantities that define the classical Kerr spacetime characterized by the rotation parameter $a = J / M$. The GUP-corrected metric can be found by replacing the BH mass with
\begin{equation}
M \rightarrow M\left(1+\beta_0\right),  
\end{equation}
where $\beta_0$ is the GUP parameter. Moreover, the GUP-corrected spacetime is obtained by replacing
\begin{equation}
\begin{aligned}
r_S &\rightarrow 2M\left(1+\beta_0 \right)=2M\gamma^{-1},\\
a &\rightarrow a\left(1+\beta_0 \right)^{-1}=a\gamma, \nonumber
\\
\rho^2 &\rightarrow r^2 + a^2 \gamma^2 \cos^2\theta,\\
\Delta &\rightarrow r^2 - 2M\gamma^{-1}r + a^2\gamma^2,
\end{aligned}
\end{equation}
where $\gamma = \left(1 +\beta_0 \right)^{-1}$.

In order to determine the limits on the GUP parameter $\beta_0$, we solve numerically the null geodesic equations in strong gravity conditions for different values of $\beta_0$; then, we perform the OAM beam analysis to relate the OAM spectrum to the rotation of the GUP-modified BHs as it was made in Ref. \cite{29} to measure the rotation parameter of the compact object in M87 from EHT data. Analysing the OAM content in the lensed light observed in the neighborhoods of the black hole with the spiral spectrum \cite{OAMspectrum}, the rotation can be obtained by considering the ratio $q$ between the $\ell =1$ and $\ell = -1$ OAM spectral components \cite{28b,29}.

\begin{figure}
\begin{center}
\includegraphics[width=4.2cm]{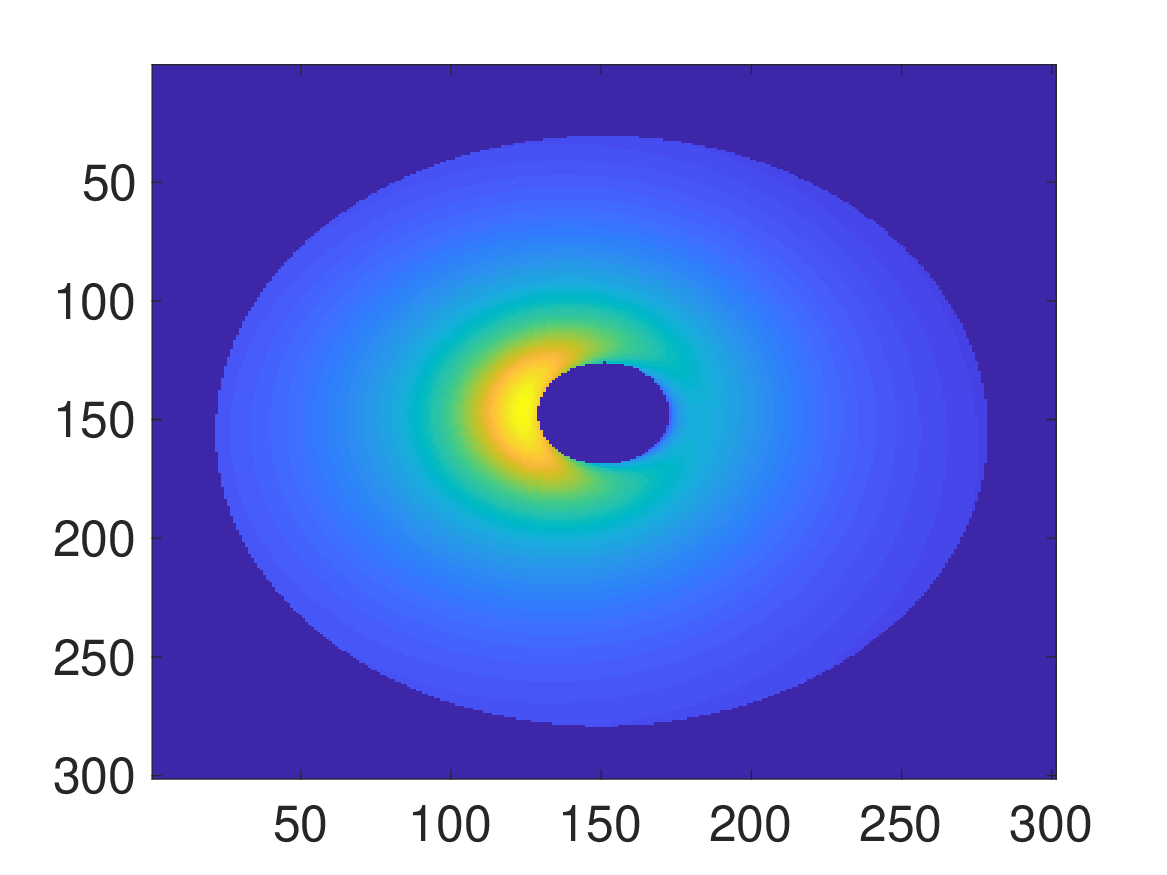}
\includegraphics[width=4.2cm]{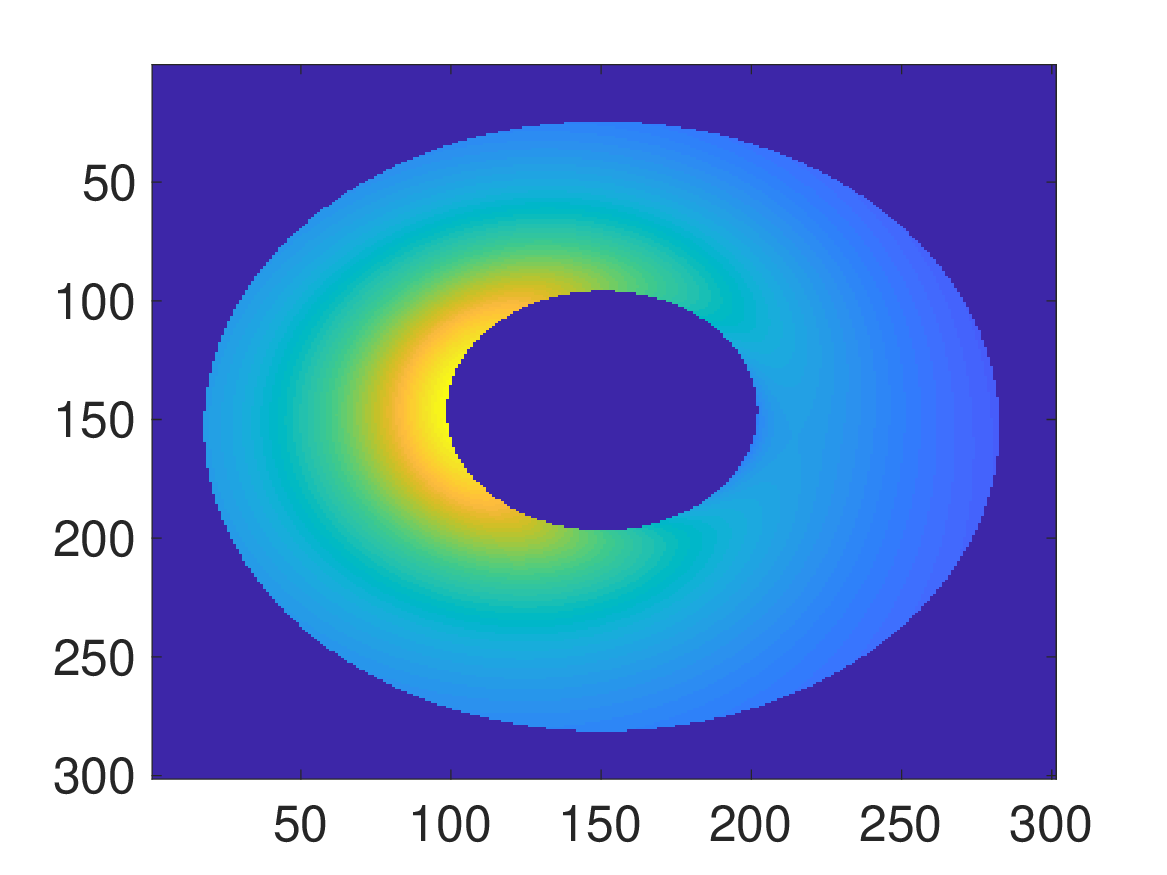}
\includegraphics[width=4.2cm]{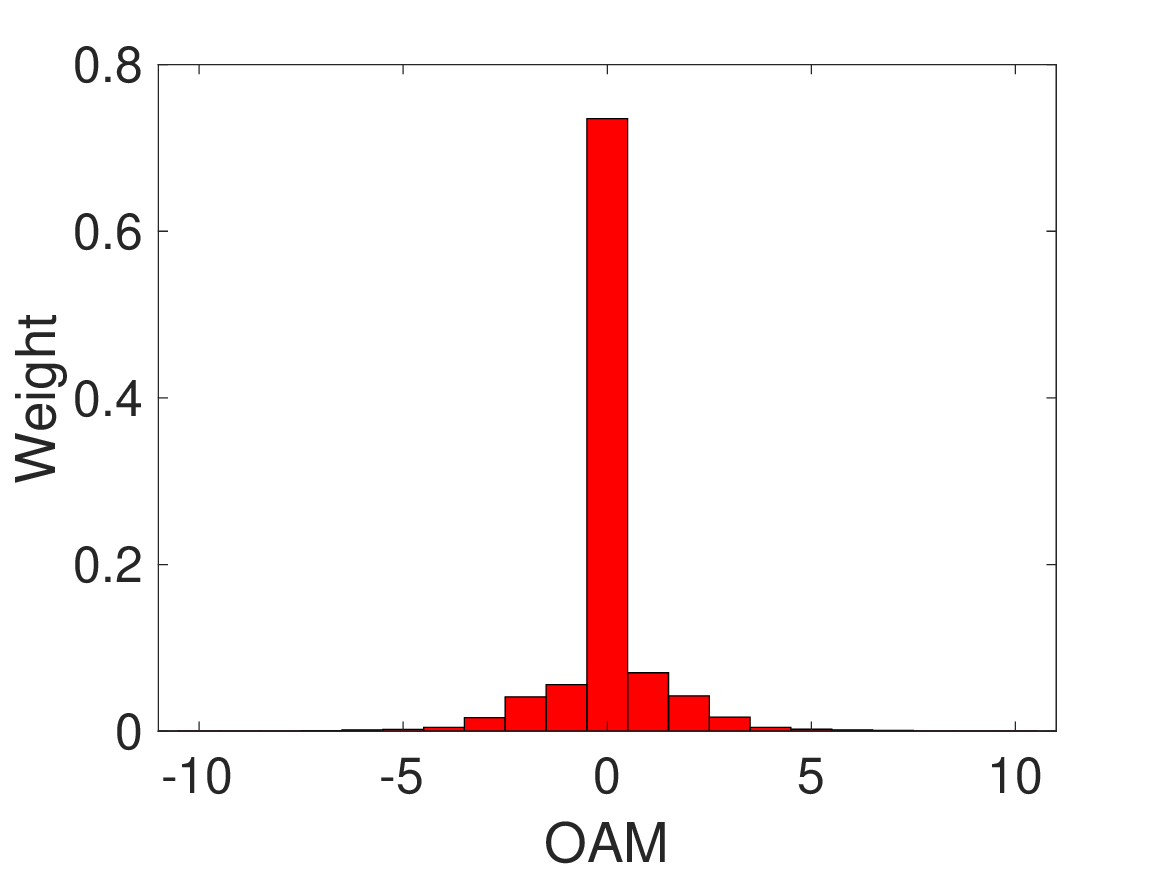}
\includegraphics[width=4.2cm]{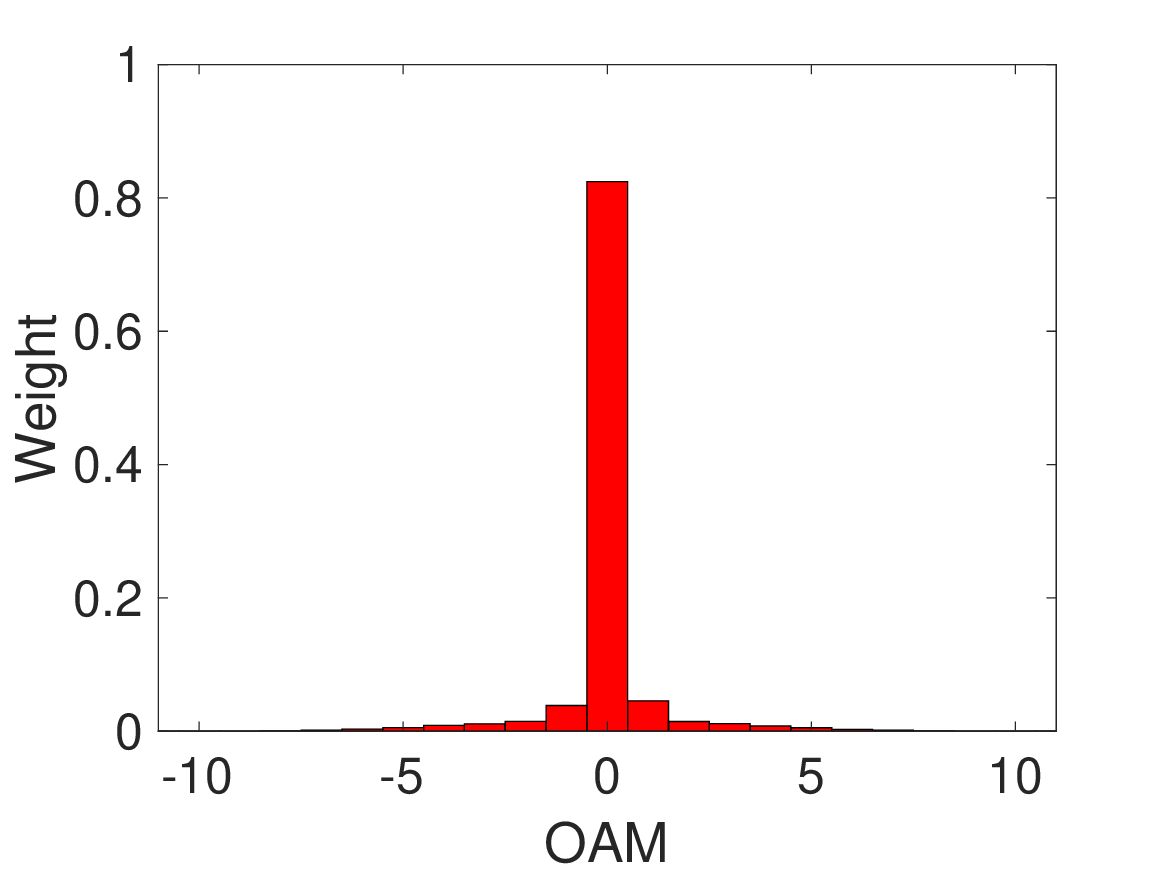}
\end{center}
\caption{\textbf{Upper panels:} Intensity of the z-component of an ideal accretion disk around a rotating BH with rotation parameter $a=0.85$ for a Kerr with $\beta_0=0$ (left), and a GUP BH with $\beta_0=1$ (right). The scales are in arbitrary numbers. We note that the more the GUP parameter grows the more the shadow of the BH increases.
\\
\textbf{Lower panels:} the corresponding OAM spectra for Kerr ($\beta_0=0$) and GUP BH ($\beta_0=1$) show that the rotation of a GUP BH imposes a smaller twist to the lensed light as the parameter $\beta_0$ increases (see text).}
\label{fig1}
\end{figure}

To determine the effects of the GUP-corrected gravitational field, we fixed the size of the accretion disk (AD) to $r_{disk}=10$, given in units of BH masses of an ideal Kerr BH, and the rotation parameter $a=0.85$ with inclination $i=17^\circ$. It is immediately evident, as seen in the upper panels of Fig. \ref{fig1}, that the result of the gravitational lensing due to a rotating compact object in the presence of gravitational GUP corrections appears different from that of a Kerr BH described by the standard equations of GR. The image in the inner parts of the AD nearby the BH results modified, as expected \cite{Neves,Jusufi}.
In this way, a double action on the vorticity of the lensed EM radiation occurs: the central region of the AD, where the gravitational field of the BH is more effective in imprinting its swirling effects to the lensed light, is replaced by regions of the GUP-corrected spacetime that are  less efficient at imprinting such a twist.
At the same time, the size of the GUP BH increases, invading the center of the AD and modifying the local geometry of the spacetime and the inner shape of the AD itself. This is evident from the spiral spectra in the lower panels of Fig. \ref{fig1}: the standard Kerr solution shows an OAM spectrum richer in high-order components with $q=1.2549$ and a central peak with height $h_{\ell=0}=0.7350$. On the other side, the GUP spacetime shows $q=1.1842$ and a higher $\ell=0$ component ($h_{\ell=0}=0.8243$), indicating a less effective transfer of OAM.

\begin{figure}
\begin{center}
\includegraphics[width=9.3cm]{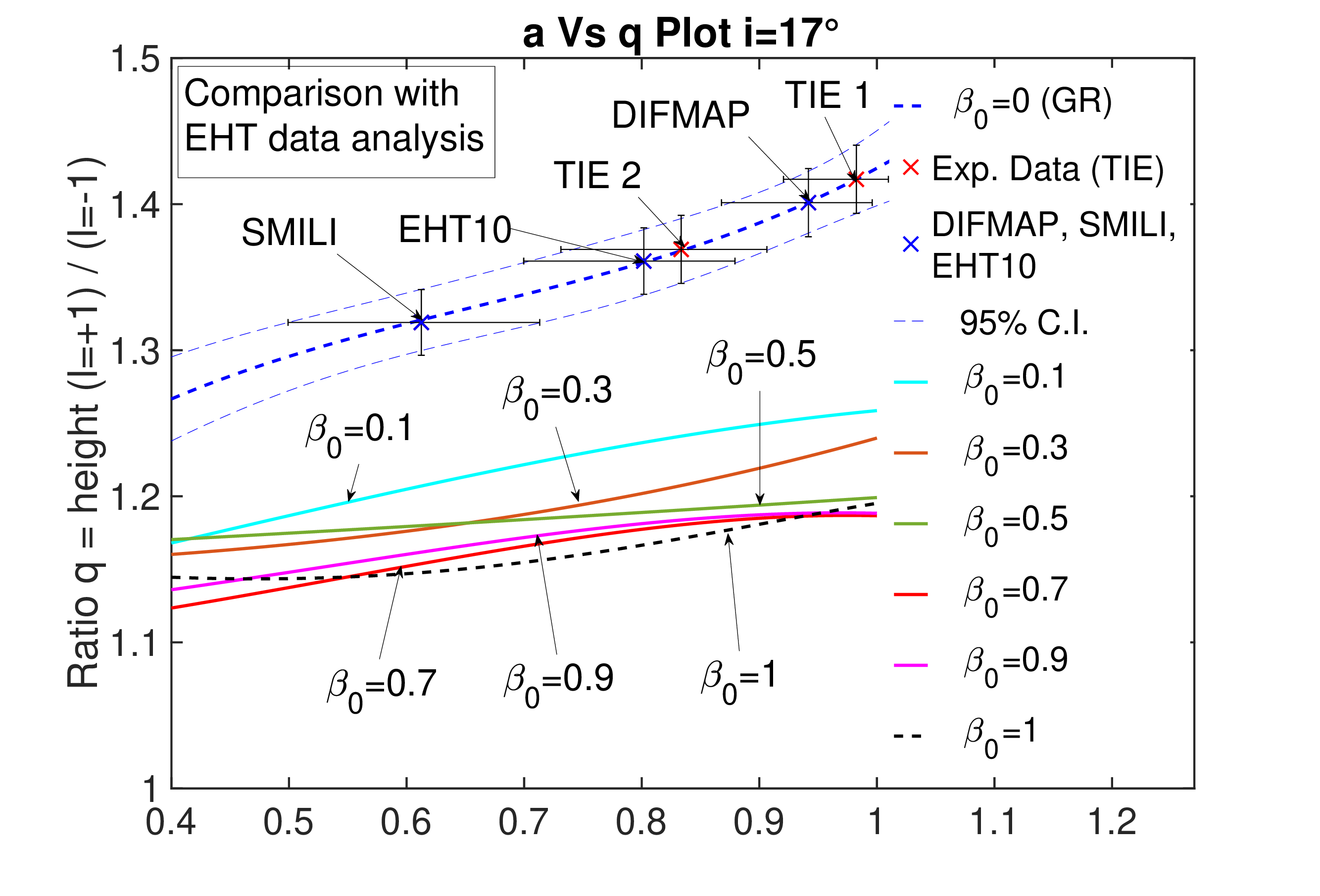}
\end{center}
\caption{Plot of the BH rotation parameter $a$ Vs $q$ (the asymmetry parameter of the  OAM spectrum) of rotating black holes in GUP geometries, obtained for different values of the rescaled GUP parameter $\beta_0$. The numerical results have been obtained with KERTAP and the curves between the estimates have been interpolated with Hilbert polynomials.
Then, the results of the numerical simulations have been compared with those obtained from the numerical simulations of Kerr metric and the analysis of the data performed by the EHT team in their observational epochs 1 and 2. The parameter $q$ decreases as $\beta_0$ increases: GUP BHs with high $\beta_0$ values twist much less the light lensed by them.}
\label{fig2}
\end{figure}

In Fig. \ref{fig2} we report the simulations of the rotation parameter $a$ Vs $q$ (OAM asymmetry parameter) of Kerr and GUP BHs with rotation parameters varying in the interval $0.4 \leq a \leq 0.985$ and GUP parameter $0 < \beta_0 \leq 1$.
The more the GUP parameter $\beta_0$ increases, the more the corresponding $(a,q)$ curve is confined to lower regions of the plot, towards values of the parameter $q$. This effect clearly indicates that, due to the GUP corrections, the rotation of the compact object is less effective in the transfer of OAM to the lensed light. By comparing the BH parameters obtained with other independent methods or from experimental data, one can constrain the values of the GUP parameter with twisted light.
To this aim, we compare our numerical results with those obtained by the EHT collaboration \cite{30a,30b,30c,30d,30e,30f} in epochs 1 and 2 (TIE 1 and 2) \cite{29} as reported in Fig. \ref{fig2}, and with the simulations of a Kerr BH with the parameters of M87*.
In this case, we assume for M87* a rotation parameter $a = 0.90\pm0.05$ ($95\%$ confidence level) and inclination $i = 17^\circ \pm 2^\circ$, according to Ref. \cite{29}. 
To determine the limits on $\beta_0$ we take as reference the $ 95 \% $ confidence zone defined by the analysis of M87* OAM data, and we find from the numerical simulations that  the GUP parameter is restricted in the interval $0 < \beta_0  \leq 0.01064$. 
This shows that the use of the additional information encoded in the phase of OAM beams allows us to extract more information from the experimental data, obtaining a better upper limit to the value of the GUP parameter $\beta_0$, smaller than two up to three orders of magnitude than the one previously estimated with rotating BHs. We did not set as lower limit $\beta_0=0$ in the inequality because the errors introduced by the numerical simulations are of the order of $5 \times 10^{- 7}$, and the experimental errors introduced by the EHT OAM data analysis are $\sim 2 \% $, setting the indetermination on the GUP parameter to $2.128 \times 10^{- 4}$.
The only way to improve this result is to use full detailed numerical simulations of M87*, as performed by EHT, coupled with the OAM analysis of the experimental data.

\vspace{-0.2cm}

\section{Conclusion remarks}
In this letter we have determined a new upper limit on the rescaled GUP parameter $\beta_0=\beta / 2M^2$ thanks to the OAM analysis of the twisted light from the compact object observed in M87 \cite{28b,29} by the Event Horizon Telescope collaboration \cite{30a,30b,30c,30d,30e,30f}. By using the additional information encoded in the phase of the OAM beam, we found that $\beta_0  \leq 0.01064 \pm 0.0002128$, a more stringent bound with respect to the one recently obtained in \cite{Jusufi}, where a black hole shadow analysis has been performed.
This shows how the novel technique described here could become a new important tool for astronomy and astrophysics.
\vspace{-0.3cm}
\section*{Acknowledgements}The authors thank Dennis Durairaj for his suggestions.

\end{document}